\documentclass[10pt,aps,pra,twocolumn,superscriptaddress,floatfix]{revtex4-1}

\usepackage{amsmath}
\usepackage{xcolor,soul}
\usepackage{graphicx}

\begin{document}


\title{Photoionization cross sections of the 5S${}_{1/2}$ and 5P${}_{3/2}$ states of Rb in simultaneous magneto-optical trapping of Rb and Hg}



\author{M. Witkowski}
\email[]{mwitkowski@uni.opole.pl}
\affiliation{Institute of Physics, University of Opole, Oleska 48, PL-45-052  Opole, Poland}
\affiliation{Institute of Physics, Faculty of Physics, Astronomy and Informatics, Nicolaus Copernicus University, Grudzi\c{a}dzka 5, PL-87-100 Toru\'n, Poland}
\author{R.~Munoz-Rodriguez}
\affiliation{Institute of Physics, Faculty of Physics, Astronomy and Informatics, Nicolaus Copernicus University, Grudzi\c{a}dzka 5, PL-87-100 Toru\'n, Poland}
\author{A.~Raczy\'nski}
\affiliation{Institute of Physics, Faculty of Physics, Astronomy and Informatics, Nicolaus Copernicus University, Grudzi\c{a}dzka 5, PL-87-100 Toru\'n, Poland}
\author{J.~Zaremba}
\affiliation{Institute of Physics, Faculty of Physics, Astronomy and Informatics, Nicolaus Copernicus University, Grudzi\c{a}dzka 5, PL-87-100 Toru\'n, Poland}
\author{B.~Nag\'orny}
\affiliation{Institute of Physics, Faculty of Physics, Astronomy and Informatics, Nicolaus Copernicus University, Grudzi\c{a}dzka 5, PL-87-100 Toru\'n, Poland}
\affiliation{Time and Frequency Department, Astrogeodynamic Observatory of Space    Research Center, Borowiec, Drapa\l{}ka 4, PL-62-035 K\'ornik,  Poland}
\author{P.~S.~\.Zuchowski}
\affiliation{Institute of Physics, Faculty of Physics, Astronomy and Informatics, Nicolaus Copernicus University, Grudzi\c{a}dzka 5, PL-87-100 Toru\'n, Poland}
\author{R. Ciury\l{}o}
\affiliation{Institute of Physics, Faculty of Physics, Astronomy and Informatics, Nicolaus Copernicus University, Grudzi\c{a}dzka 5, PL-87-100 Toru\'n, Poland}
\author{M.~Zawada}
\affiliation{Institute of Physics, Faculty of Physics, Astronomy and Informatics, Nicolaus Copernicus University, Grudzi\c{a}dzka 5, PL-87-100 Toru\'n, Poland}


\date{\today}

\begin{abstract}
  We report the measurement of the photoionization cross sections of the 5S${}_{1/2}$ and 5P${}_{3/2}$ states of ${}^{87}$Rb in a two-species Hg and Rb magneto-optical trap (MOT) by the cooling laser for Hg. The photoionization cross sections of Rb in the  5S${}_{1/2}$ and 5P${}_{3/2}$ states at 253.7~nm are determined to be $1^{+4.3}_{-1}\times10^{-20}~\text{cm}^2$ and $4.63(30)\times 10^{-18}\text{cm}^2$, respectively. To measure the 5S${}_{1/2}$ and 5P${}_{3/2}$ states fractions in the MOT we detected photoionization rate of the 5P${}_{3/2}$ state by an additional 401.5~nm laser. The photoionization cross section of Rb in the 5P${}_{3/2}$ state at 401.5~nm is determined to be $\text{1.18(10)}\times10^{-17}~\text{cm}^2$.

\end{abstract}


\maketitle

\section{Introduction}

Photoionization of atoms is a known source of losses in multi-species magneto-optical trapping when photons from the cooling beam for one species are energetic enough to ionize excited state of the other species~\cite{Hensler2004,Aoki2013}. In this article, we discuss the ionization process in the two-species Hg-Rb MOT. Simultaneous trapping of cold Hg and Rb opens the way toward experiments in the field of fundamental physics. Particularly interesting features of Hg such as large mass, high nuclear charge, and variety of isotopes make this element relevant for experimental tests of Standard Model~\cite{Flambaum2009,Angstmann2004,Derevianko2014,Wcislo2016,Bouchiat1977,Delaunay2016a}. Moreover, due to its low sensitivity for black-body radiation, Hg is regarded as the perfect candidate for the most accurate optical atomic clock~\cite{Hachisu2008,Yi2011}. On the other hand, HgRb molecule is appropriate for the search for the electric dipole moment of the electron~\cite{Meyer2009,Borkowski2017}. 

For the future design of co-trapping experiments in Hg and Rb mixture it is critical to address the problem of losses of latter atom due to UV radiation, which originates from cooling laser for Hg atoms.  Earlier predictions of Aymar \cite{Aymar1984} suggest that it is likely that the Cooper minimum \cite{Fano1968} for the photoionization of Rb should be observed close to Hg cooling transition. The main goal of this paper is to elucidate previous theoretical predictions and estimate the influence of the 253.7~nm light on stability of the Rb MOT. 

The wavelength of the cooling transition is 253.7~nm for Hg which means the Hg cooling light can ionize Rb in both ground 5S${}_{1/2}$ and excited 5P${}_{3/2}$  states since photoionization from these states requires photon wavelength shorter than 297~nm and 479~nm, respectively. The resulting losses restrict the number of Rb atoms in Hg-Rb MOT. However, the respective cross sections for photoionization are small enough to allow simultaneous trapping of both species. Moreover, one can determine the values of the photoionization cross sections from observed atom losses in Hg-Rb MOT setup.

The measurement is based on the method reported in~\cite{Dinneen1992} where atoms confined within a MOT were ionized from excited state by an additional laser radiation and a subsequent decay of the trap fluorescence was measured. To date, several measurements of the cross sections of excited P and D states of various elements~\cite{Gabbanini1997,Patterson1999,Marago1998,Duncan2001,Wippel2001,Claessens2006,Ciampini2002} were reported for various wavelengths. It was also observed that the cooling photons in Cd MOT can ionize Cd from the excited ${}^{1}$P${}_{1}$ state of the cooling transition~\cite{Brickman2007}.

In the measurements reported here, since the 253.7~nm laser UV light ionizes Rb from both ground and excited states, relative populations in these states have to be determined to derive the photoionization cross sections. The excited 5P${}_{3/2}$ state fraction was measured by a photoionization 
by an additional 401.5~nm laser.

\section{Photoionization in a dual magneto-optical trap}
 In this paper, we study experimental sequences where only Rb atoms are trapped in the Hg-Rb MOT setup.
The rate equation for the number of Rb atoms $N_{Rb}(t)$ in the Hg-Rb MOT setup can be written as~\cite{Gabbanini1997}

\begin{equation}
\begin{split}
\frac{d N_{Rb}}{dt} = L_{Rb}(I^{253.7}_P)- \left(\gamma_{Rb} + \gamma^{253.7}_P\right) N_{Rb} \\- \beta_{RbRb}\int dr^3 n^2_{Rb}, \label{eq:rate_full}
\end{split}
\end{equation}
\noindent
where $L_{Rb}(I^{253.7}_P)$ is the Rb trap loading rate in the presence of ionizing radiation, $\gamma_{Rb}$ is the loss coefficient due to collisions with background gases, $\gamma^{253.7}_P$ is the loss coefficient due to photoionization by 253.7~nm light, $\beta_{RbRb}$ is the loss coefficient due to the light-assisted collisions between the Rb atoms, and $n_{Rb}$ is the spatial  density of Rb.
In general, the photoionizing radiation of intensity $I^{253.7}_P$ modifies the trap loading rate $L_{Rb}$, as it removes a fraction of Rb atoms to be captured~\cite{Marago1998}.
However, at fixed wavelength and intensity the trap loading rate $L_{Rb}$ can be treated as constant.

The collisions in the equation~\eqref{eq:rate_full} between cold Rb atoms can be ignored if the density in the Rb MOT is made low enough, e.g.~by decreasing the intensity of the trapping beams. 
 It yields  the rate equation


%
%
%
%

\begin{equation}
\frac{d N_{Rb}}{dt} = L_{Rb}(I^{253.7}_P) - \left(\gamma_{Rb} + \gamma^{253.7}_P\right) N_{Rb}.  \label{eq:rate_unRb}
\end{equation}

\noindent
The above equation can be integrated with the initial condition $N_{Rb}(t=0)=0$ and the final formula for $N_{Rb}(t)$ reads

\begin{equation}
N_{Rb}(t) = \frac{L_{Rb}(I^{253.7}_P)}{\gamma_{Rb} + \gamma^{253.7}_P}\left[1-\exp\left(-\left(\gamma_{Rb} + \gamma^{253.7}_P\right) t \right)\right] \label{eq:sol_loadRb}.
\end{equation}

\begin{figure}
 \includegraphics[width=\columnwidth]{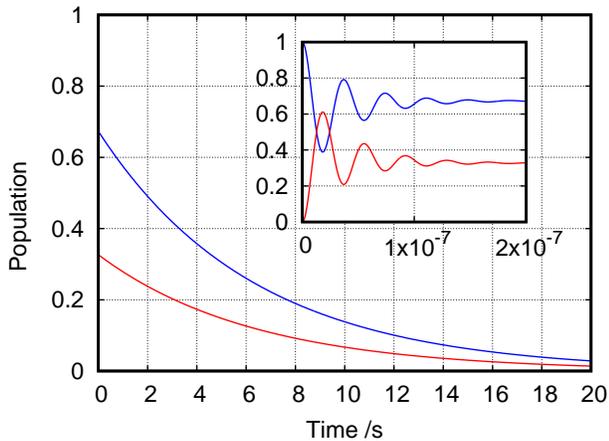}
 \caption{Results of numerical simulations of the dynamics of the rubidium atom in the presence of the 253.7~nm ionization radiation. The simulations have been performed for the typical initial excited state fraction equals to 0.32. Both ground (blue) and excited state (red) populations exhibit  a perfectly exponential decay.   Rabi oscillations before reaching  the steady state are shown in the inset. \label{fig:num_sim}}
 \end{figure}

If we assume that the photoionization processes from ground and excited states are independent, the loss coefficient $\gamma^{253.7}_P$ due to photoionization by 253.7~nm photons can be expressed as~\cite{Ciampini2002}

\begin{equation}
\gamma^{253.7}_P  =\left( \rho_{5S_{1/2}} \sigma^{253.7}_{5S_{1/2}} + \rho_{5P_{3/2}}\sigma^{253.7}_{5P_{3/2}}\right)\frac{I^{253.7}_P}{h\nu^{253.7}_P}, \label{eq:gamma253}
\end{equation}

\noindent
where $\rho_{5S_{1/2}}$ and $\rho_{5P_{3/2}}$ are the population fractions of the $5S_{1/2}$ and $5P_{3/2}$ states, respectively (it is assumed that $\rho_{5S_{1/2}}+\rho_{5P_{3/2}} = 1$ in the Rb MOT), $I^{253.7}_P$ is the photoionizing laser intensity, $h \nu_P^{253.7}$ is its photon energy,   $\sigma^{253.7}_{5S_{1/2}}$ and $\sigma^{253.7}_{5P_{3/2}}$ are the photoionization cross sections from these states. 
{Since the validity of the latter assumption is crucial for further investigation, even if this assumption is generally accepted, we have carefully examined the independence of both contributions. 

To analyze  the behavior  of the rubidium atom exposed to ionization radiation we have represented it by a two-level model in the rotating wave approximation including the states $5S_{1/2}$ and $5P_{3/2}$ and we have solved the corresponding Bloch equations (\ref{eq:bloch})

\begin{widetext}
\begin{eqnarray}
\frac{d}{dt}
\left(
\begin{array}{c}
\rho_{5S_{1/2}}\\\rho_{5P_{3/2}}\\s\\s^*\end{array}\right)=
\left(\begin{array}{cccc}-i\Gamma_{5S_{1/2}}^{253.7}&i\Gamma_{sp}&U&-U\\
0&-i\Gamma_{5P_{3/2}}^{253.7}-i\Gamma_{sp}&-U&U\\
U&-U&-\Delta-i\Gamma/2&0\\
-U&U&0&\Delta-i\Gamma/2
\end{array}\right)
\left(
\begin{array}{c}
\rho_{5S_{1/2}}\\\rho_{5P_{3/2}}\\s\\s^*\end{array}\right).
 \label{eq:bloch}
\end{eqnarray}

\end{widetext}

\noindent
In the above equations $\rho_{5S_{1/2}}$ and $\rho_{5P_{3/2}}$ are the diagonal elements of the density matrix, $s\equiv\langle5P_{3/2}|\rho|5S_{1/2}\rangle$ is its nondiagonal element (coherence), $\Gamma_{sp}$ is the relaxation rate due to spontaneous emission from the upper state to the lower one, $\Gamma_{S_{1/2}}^{253.7}$ and $\Gamma_{P_{3/2}}^{253.7}$ are ionization rates from the states $5S_{1/2}$ and $5P_{3/2}$, respectively, due to the 253.7~nm ionizing laser field. The relaxation rate $\Gamma/2$ for the coherence between the two states is $(\Gamma_{sp}+\Gamma_{S_{1/2}}^{253.7}+\Gamma_{P_{3/2}}^{253.7})/2$~\cite{Tannoudji1992}. 
The coupling between  $5S_{1/2}$ and $5P_{3/2}$ states due to the trapping laser field detuned by $\Delta=\omega-\omega_0$ from the atomic resonance $\omega_0=(E_{5P_{3/2}}-E_{5S_{1/2}})/\hbar$ is characterized by the Rabi frequency U. 

In the absence of the ionizing laser, i.e., for the ionization rates $\Gamma_{S_{1/2}}^{253.7}=\Gamma_{P_{3/2}}^{253.7}=0$, the evolution ends up in a steady state after a few lifetimes $1/\Gamma_{sp}$.  The parameters of the steady state (populations of the two states and the corresponding coherence) depend on the model parameters $(U,\Delta,\Gamma_{sp})$. In the steady state the asymptotic values are

\begin{eqnarray}
\rho^{steady}_{5S_{1/2}}=\frac{U^2+\Delta^2+\Gamma_{sp}^2/4}{2U^2+\Delta^2+\Gamma_{sp}^2/4},\nonumber\\
\rho^{steady}_{5P_{3/2}}=\frac{U^2}{2U^2+\Delta^2+\Gamma_{sp}^2/4},\\
s^{steady}=\frac{U(\Delta-i\Gamma_{sp}/2)}{2U^2+\Delta^2+\Gamma_{sp}^2/4}.\nonumber
\end{eqnarray}

\noindent
Switching the ionizing laser on results in a slow depopulation of both $5S_{1/2}$ and $5P_{3/2}$ states. 

We have solved the Bloch equations (\ref{eq:bloch}) using the Laplace transform. The result is that the populations and coherence as functions of time are superpositions of terms of the form $\exp(-iz_jt)$, where $z_j
$ are the eigenvalues of the matrix of the Bloch equations. Note that in the absence of ionization one of the eigenvalues is zero and its corresponding term  is the only one which survives after the steady state has been established. An example of the time dependence of the population of both states calculated for $\rho_{5P_{3/2}}(t=0)=0.32$, $\Gamma_{sp}=2\pi \times 6.06$~MHz~\cite{Volz1996}, $\Delta=2\Gamma_{sp}$ is shown in Fig.~\ref{fig:num_sim}. Except for a relatively short  time  interval, where the Rabi oscillations appear (inset in Fig.~\ref{fig:num_sim}), both states are depopulated exponentially.

The term of the matrix that includes the ionization rates can be treated as a perturbation
and evaluated using the perturbation theory. The perturbation matrix is diagonal with the matrix elements: $(-i\Gamma_S^{253.7}$, $-i\Gamma_P^{253.7}$, $-i(\Gamma_S^{253.7}+\Gamma_P^{253.7})/2$,
$-i(\Gamma_S^{253.7}+\Gamma_P^{253.7})/2)$.
Because the unperturbed matrix is  non-Hermitian, applying  the perturbation theory requires distinguishing between right and left eigenvectors.
We are only interested in the correction to the zero eigenvalue. The corresponding right and left eigenvectors (unnormalized) are, respectively,  $\psi=\left(U^2+\Delta^2+\Gamma^2/4,U^2,U(\Delta-i\Gamma/2),U(\Delta+i\Gamma/2)\right)$ and $\tilde{\psi}=(1,1,0,0)$.
The corrected eigenvalue is

\begin{equation}
\gamma_{ph}\equiv\frac{\langle\tilde{\psi}|W|\psi\rangle}{\langle\tilde{\psi}|\psi\rangle}
=\rho^{steady}_{5S_{1/2}}\Gamma_{5S_{1/2}}^{253.7}+\rho^{steady}_{5P_{3/2}}\Gamma_{5P_{3/2}}^{253.7}.\label{eq:corr_eigen}
\end{equation}

\noindent
The effective ionization rate, common for both levels, is a mean value of the two ionization rates weighted with the steady state populations. Since the cross sections are proportional to the probabilities, the Eq. \ref{eq:corr_eigen} justifies the Eq. \ref{eq:gamma253}.
}

The excited 5P${}_{3/2}$ state fraction $\rho_{5P_{3/2}}$ can be detected with the photoionization by a light at a wavelength between 297~nm and 479~nm. The solution of loading rate equation of the weak Rb MOT in a presence of the 401.5~nm light is 
analogously given as

\begin{equation}
N(t) = \frac{L_{Rb}(I^{401.5}_P)}{\gamma_{Rb} + \gamma^{401.5}_P}\left[1-\exp\left(-\left(\gamma_{Rb} + \gamma^{401.5}_P\right) t \right)\right],  \label{eq:rate_unRb401}
\end{equation}

\noindent
where

\begin{equation}
\gamma^{401.5}_P  = \rho_{5P_{3/2}}\sigma^{401.5}_{5P_{3/2}} \frac{I^{401.5}_P}{h\nu^{401.5}_P}\label{eq:gamma401}
\end{equation}

\noindent
is the loss coefficient due to photoionization by a 401.5~nm light, $\sigma^{401.5}_{5P_{3/2}}$ is the photoionization cross section, $I^{401.5}_P$ is the photoionizing laser intensity, and $h\nu^{401.5}_P$ is its photon energy.

\section{Measurement of the photoionization rate at 401.5 \MakeLowercase{nm} and 253.7 \MakeLowercase{nm}}

Our Hg-Rb MOT setup has been described in detail elsewhere~\cite{Witkowski2017}, so only its most essential elements  are presented below.

For cooling and repumping the ${}^{87}$Rb commercial external-cavity diode lasers (ECDL) are used. The cooling laser beam is additionally amplified by a fibre-coupled tapered amplifier up to 700~mW.  
The frequency of these laser beams are stabilized by the Pound-Drever-Hall (PDH)~\cite{Hansch1980} locking and the saturated absorption spectroscopy to an atomic resonance in a rubidium vapour
cell. 

Photoionization of ground state Rb atoms is induced by the same laser system which is used to cool and trap Hg atoms.
In this setup a 1014.8~nm laser light is used as a fundamental source and its light is frequency-quadrupled in a system of two doubling stages.
The source laser is composed of a home-made ECDL and a
tapered amplifier. Power build-up bow-tie cavities are used to improve the efficiency of frequency doubling on a LBO and BBO crystals. 
The Hg cooling laser frequency is stabilized 
to the narrow (7.5 kHz) ${}^1$S${}_0$ -- ${}^3$P${}_1$ transition in Sr by
a stability transfer from an ultra-stable laser~\cite{Lisak2012,Cygan2013,Morzynski2015}. 
A small part of the fundamental 1014.8~nm light is uncoupled from the main beam
before quadrupling and provided to a transfer cavity. The fundamental  laser is locked to the ultra-stable
laser by the PDH method. The same method is used to stabilize the length of the transfer cavity to the ultra-stable
laser.

The vacuum system  consists of two perpendicular arms connected to the science chamber. These
two arms provide beams of mercury and rubidium atoms, respectively. The rubidium atomic beam is produced by a recirculating oven and slowed in the Zeeman slower.
Mercury has a high vapour pressure at room temperature and  it needs to be cooled to maintain
high-vacuum conditions. To lower the temperature of mercury a two-stage thermoelectric cooler is used. 
Between the cooler and the science chamber a two-dimensional magneto-optical trap (2D-MOT) is placed. 
This stage, although not required, since a 3D-MOT can be loaded directly from
the mercury background vapour, improves the loading rate substantially.

The characterization of the MOT can be performed either by analysis of fluorescence images
of the MOT taken by a CCD camera or by measuring the fluorescence light from the Rb and Hg atoms by
 a calibrated photodiode and a photomultiplier tube, respectively.

\begin{figure}
 \includegraphics[width=\columnwidth]{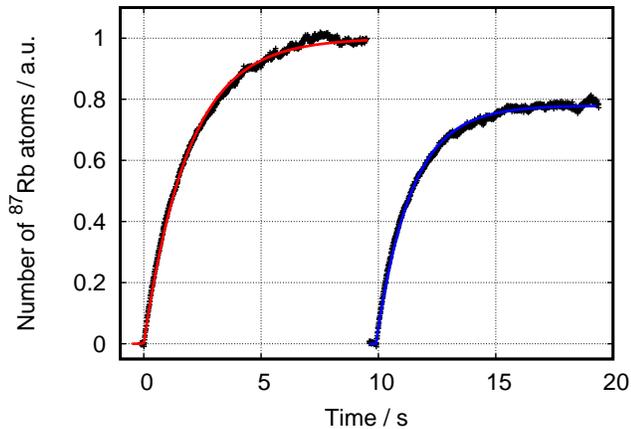}%
 \caption{Loading curves of the ${}^{87}$Rb MOT  in the presence  (right, blue curve) and the absence (left, red curve) of the Hg cooling beams. Total intensity of the cooling light is 35~mW/cm${}^2$ and 45.3~mW/cm${}^2$ for 253.7~nm and 780~nm wavelength, respectively. Both curves are the best fits given by the equation~(\ref{eq:sol_loadRb}).\label{fig:single}}
 \end{figure}

 \begin{figure}
 \includegraphics[width=\columnwidth]{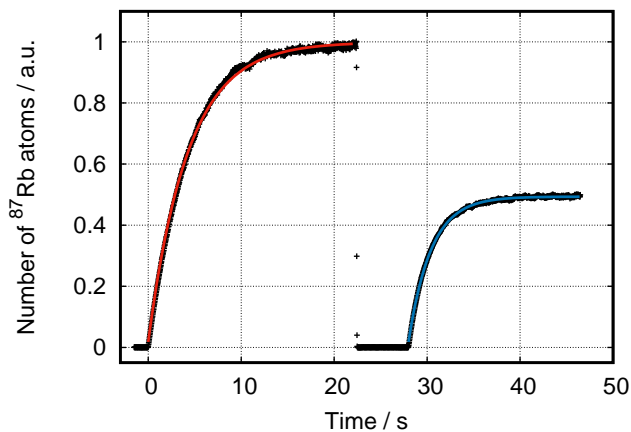}%
 \caption{Loading curves of the ${}^{87}$Rb MOT  in the presence  (right, blue curve) and the absence (left, red curve) of 401.5~nm light with $I^{401.5}_P$=30~mW/cm${}^2$. Both curves are the best fits given by the equation~\eqref{eq:rate_unRb401}.\label{fig:load}}
 \end{figure}

Fig.~\ref{fig:single} shows the loading curves of the ${}^{87}$Rb MOT  in the presence and the absence of the Hg MOT cooling beams. 
To quantify how ionizing radiation depletes the Rb trap the source of Hg atoms is blocked. 
The losses introduced by  the 253.7 laser UV light photoionization are clearly visible. The evolution of the number of Rb atoms is governed by the equation~\eqref{eq:rate_unRb}. To characterize the UV light photoionization losses one has to know the photoionization cross sections of both ground and excited ${}^{87}$Rb states as well as the population fractions of these states (Eq.~(\ref{eq:gamma253})). 

\subsection{Calibration of ground  $5S_{1/2}$ and excited $5P_{3/2}$ states fractions}

To calibrate ground  $5S_{1/2}$ and excited $5P_{3/2}$ states fractions 
 the dependence of the photoionization rate by a light at 401.5~nm on the Rb trapping laser intensity was measured. The 401.5~nm light was generated by a free-running laser diode. 
 The rough alignment of the 401.5~nm beam was based on the geometry of the vacuum chamber whereas the fine aligning was done by maximizing atomic losses from the trap induced by photoionization.

 Typical loading curves of the ${}^{87}$Rb MOT  in the presence and the absence of the 401.5~nm light are presented in Fig.~\ref{fig:load}.
While the excited 5P${}_{3/2}$ state fraction $\rho_{5P_{3/2}}$, under standard MOT operating conditions, could be estimated from the trapping laser parameters~\citep{Townsend1995}, it can also be directly calibrated by the measurement of the saturation of the photoionization rate with trapping laser intensity~\cite{Dinneen1992}. Combining the model of the excited-state fraction of two-state atoms in the MOT

\begin{equation}
  \rho_{5P_{3/2}} = \frac{I_{780}/I_{sat}}{2I_{780}/I_{sat}+4(\Delta/\Gamma_{sp})^2+1},\label{eq:ex_calibrate}
\end{equation}

\noindent
where $I_{780}$ is the total trapping intensity, the $I_{sat}$ is the saturation intensity, $\Delta$ is the trapping laser detuning, and $\Gamma_{sp}$ is the natural decay rate of the excited state ($2\pi\times 6.06$~MHz for ${}^{87}$Rb) 
with the equation (\ref{eq:gamma401}) yields the dependence of the  loss coefficient $\gamma^{401.5}_P$ on the total trapping intensity $I_{780}$ 

\begin{equation}
  \gamma^{401.5}_P  = \sigma^{401.5}_{5P_{3/2}} \frac{I^{401.5}_P}{h\nu^{401.5}_P} \times \frac{I_{780}/I_{sat}}{2I_{780}/I_{sat}+4(\Delta/\Gamma_{sp})^2+1}.\label{eq:satcrv}
\end{equation}

The photoionization rate $\gamma_P^{401.5}$ was determined by detecting fluorescence from the MOT during its loading alternately with and without the presence of the photoionization radiation. Fig.~\ref{fig:load} shows one pair of loading curves recorded in 401.5~nm photoionization experiment. To prevent detection of unwanted stray light the radiation from the MOT was filtered by 780~nm interference filter and focused on a photodiode. To toggle the ionizing light on and off a mechanical shutter was used. 

Two consecutive loading curves were measured and fit independently according to the equation (6) giving loading rates  $\gamma_{Rb}$ and $\gamma_{Rb}+\gamma_P^{401.5}$ for ionizing light off and on, respectively. The photoionization rate $\gamma_P^{401.5}$ was determined by the difference of successive loading rates.

\begin{figure}
\includegraphics[width=\columnwidth]{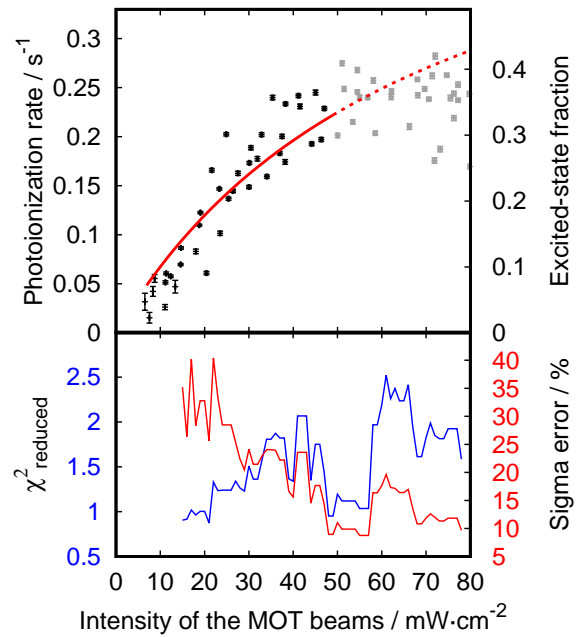}%

\caption{Top: Dependence of the 401.5~nm  ionization rate in  the ${}^{87}$Rb MOT (and the determined excited-state fraction) on the trapping laser intensity. The solid curve is a weighted fit of the equation (\ref{eq:satcrv}) with the effective saturation intensity $I_{sat}$ and $\sigma^{401.5}_{5P_{3/2}}$ being the free parameters. The dotted red line depicts its extension outside the fitting range. Experimental data are represented by scattered black and gray points, inside and outside valid fitting range, respectively.  The vertical error bars represent one standard deviation of the mean calculated for a given intensity of the MOT beams. The horizontal error bars include uncertainty associated with the fluctuations of the intensity of the MOT beams (less than 1~\%) as well as the uncertainty in the MOT size measurement (less than 1.5~\%).  The 401.5~nm ionizing laser intensity is kept at 17~mW/cm~$^2$ within 3\% of relative uncertainty.  Bottom: Dependence of the reduced chi-squared of the fit (blue) and the width of the histogram of the residuals (red) on the maximum value of the fitting range. The common minimum at 49~mW/cm${^2}$ has been chosen as the best-fit value for the dependence shown in the upper part.
\label{fig:rate_vs_I}}
 \end{figure}

Fig.~\ref{fig:rate_vs_I} depicts the measured dependence of the 401.5~nm  ionization rate in  the ${}^{87}$Rb MOT on the trapping laser intensity. The solid curve is a fit of the equation (\ref{eq:satcrv}) with the effective saturation intensity $I_{sat}$ and $\sigma^{401.5}$ being the free parameters. The determined value of $I_{sat}$ calibrate the dependence (\ref{eq:ex_calibrate}) for a given geometry and detunings of all ${}^{87}$Rb MOT beams. 

As it was shown in Ref.~\cite{Shah2007}, the equation (\ref{eq:satcrv}) fits experimental data well only for low enough intensities of the MOT beams.  To determine the optimal upper limit of the useful fitting range we have analyzed the quality of fits calculated for various ranges. We have varied the upper limit while the beginning of the fitting range was fixed at zero intensity.  The lower part of Fig.~\ref{fig:rate_vs_I} shows the dependencies of the reduced chi-squared of the fit (blue) and the width of the histogram of the residuals (red) on the maximum value of the fitting range. The general behavior of both curves is similar, except for low intensities, where the amount of points is too low to ensure reliable fit. The best-fit value corresponds to the common minimum of both curves which is equal to 49~mW/cm$^{2}$.

The appropriate range of data corresponds to a solid part of the red line in Fig.~\ref{fig:rate_vs_I}. Outside that range (dotted red line) the data points start to be scattered and the dependence of the photoionization rate on the intensity of the MOT beams is negligible making this area useless in further investigation. In comparing our experimental data with the results shown in Ref.~\cite{Shah2007}, one should note that the distribution of measured values in both sets of the data is similar  and it clearly shows a split between the scattered and the unscattered parts.

\subsection{Photoionization at 401.5~nm}

The fit of the equation (\ref{eq:satcrv}) in Fig.~\ref{fig:rate_vs_I} provides also the value of the photoionization cross section $\sigma^{401.5}_{5P_{3/2}}$. However, more precise value of $\sigma^{401.5}_{5P_{3/2}}$ is determined by measurement of the dependence of the 401.5~nm  ionization rate on the photoionizing laser intensity (Eq. (\ref{eq:gamma401})) while the intensity of the MOT beams (and consequently excited state fraction) remains constant.

To measure the average intensity of the 401.5~nm ionizing radiation seen by the atoms in the MOT we used the formula~\cite{Lowell2002} 

\begin{equation}
  \langle I \rangle =\int \int \frac{I(x,y)N(x,y)}{N_0} dx dy,\label{eq:avr_intensity}
\end{equation}

\noindent
where $N_0$ is the total number of atoms in the MOT, $I(x,y)$ and $N(x,y)$ are distributions of the intensity of the beam and the atoms, respectively. The intensity profile of the photoionizing beam was determined by a CCD beam profiler measurement to be a Gaussian function given by

\begin{equation}
  I(x,y)=I_0 exp\left( \frac{-2x^2}{\sigma_x^2} \right) exp\left( \frac{-2y^2}{\sigma_y^2} \right),\label{eq:int_distribution}
\end{equation}

\noindent
where $\sigma_x=2.096(28)$~mm and $\sigma_y=2.110(33)$~mm are $e^{-2}$ half waists of the beam. Their values and uncertainties were derived from the two-dimensional Gaussian fit. The maximum intensity $I_0$ can be expressed as 

\begin{equation}
  I_0=\frac{2PT}{\pi\sigma_x\sigma_y},\label{eq:I_0}
\end{equation}

\noindent
where $P$ is the power of the photoionizing beam and $T$ is its transmission coefficient through the fused silica window which is equal to 0.95.

The distribution $N(x,y)$ was determined by imaging fluorescence of the atoms trapped in the MOT on a CCD camera. The analysis of the image yields the Gaussian distribution

\begin{equation}
  N(x,y)=\frac{2N_0}{\pi r_x r_y}exp\left( \frac{-2x^2}{r_x^2} \right)exp\left( \frac{-2y^2}{r_y^2}\right),\label{eq:N_distribution}
\end{equation}

\noindent
where $r_x=1.47(1)$~mm and $r_y=1.54(2)$~mm are $e^{-2}$ radii of the MOT.

\begin{figure}[ht]
\includegraphics[width=\columnwidth]{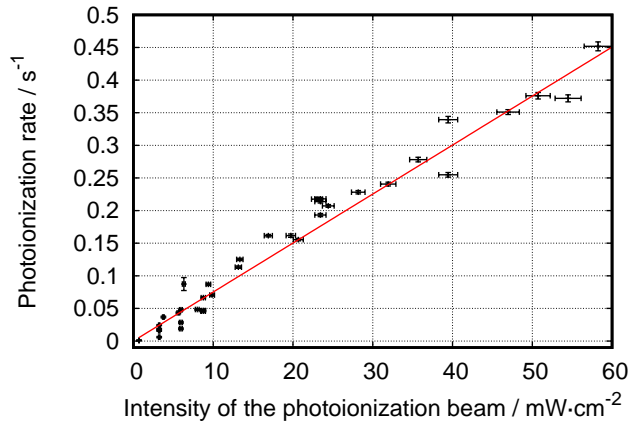}%
\caption{Dependence of the 401.5~nm photoionization  rate $\gamma^{401.5}_P$ in ${}^{87}$Rb MOT on the photoionizing laser intensity $\langle I \rangle$. The (red) solid line is a linear fit to Eq.~(\ref{eq:401lin}). The measurement was performed for the total intensity of the MOT beams equals to 35.2~mW/cm$^{2}$, i.e.\ the excited state fraction equals to 0.27.~\label{fig:401sigma_}}
\end{figure}

By putting together equations~(\ref{eq:avr_intensity})~--~(\ref{eq:N_distribution}) the formula~(\ref{eq:gamma401}) can be rewritten as

\begin{equation}
  \sigma^{401.5}_{5P_{3/2}}=\frac{h\nu_{5P_{3/2}}^{401.5}}{\rho_{5P_{3/2}}}A,\label{eq:cross_400_fit}
\end{equation}

\noindent
where $A$ is obtained 
from a weighted linear fit of the
formula

\begin{equation}
  \gamma^{401.5}_P(\langle I\rangle)=A \langle I \rangle=A I_0 \frac{\sigma_x \sigma_y}{\sqrt{\left( r_x^2+\sigma_x^2 \right)\left( r_y^2+\sigma_y^2 \right)}}\label{eq:401lin}
\end{equation}

\noindent
to the data depicted in Fig.~\ref{fig:401sigma_}. 

We determined from equations~(\ref{eq:cross_400_fit}) and (\ref{eq:401lin}) that the value of the photoionization cross section at 401.5~nm is 

\begin{equation}
\sigma^{401.5}_{5P_{3/2}}=1.18(10)\times 10^{-17}~\text{cm}^{2}. 
\end{equation}

\noindent
The uncertainty $u(\sigma^{401.5}_{5P_{3/2}})$ has been estimated using standard uncertainty analysis

\begin{equation}
  u(\sigma^{401.5}_{5P_{3/2}})=h\nu_{5P_{3/2}}^{401.5} \sqrt { \left( \frac{u(A)}{ \rho_{5P_{3/2}}}\right)^2+\left(\frac{Au( \rho_{5P_{3/2}}) }{\rho_{5P_{3/2}}^2} \right)^2},\label{eq:uncert} 
\end{equation}

\noindent 
where $u(A)$ is the uncertainty of $A$ from the weighted linear fit of Eq. (\ref{eq:401lin}).  The uncertainty $u(A)$ involves the accuracy of the  measurements of the Rb MOT radii, the waist, and the intensity of the ionization beam. The value of the $u(\rho_{5P_{3/2}})$ has been extracted from the fit of the equation~(\ref{eq:satcrv}) (solid red line in Fig.~\ref{fig:rate_vs_I}). Both terms in formula~(\ref{eq:uncert}) contribute to the total uncertainty $u(\sigma^{401.5}_{5P_{3/2}})$ with comparable level of significance.

\subsection{Photoionization at 253.7~nm}

\subsubsection{$5S_{1/2}$ state}

\begin{figure}[ht]
 \includegraphics[width=\columnwidth]{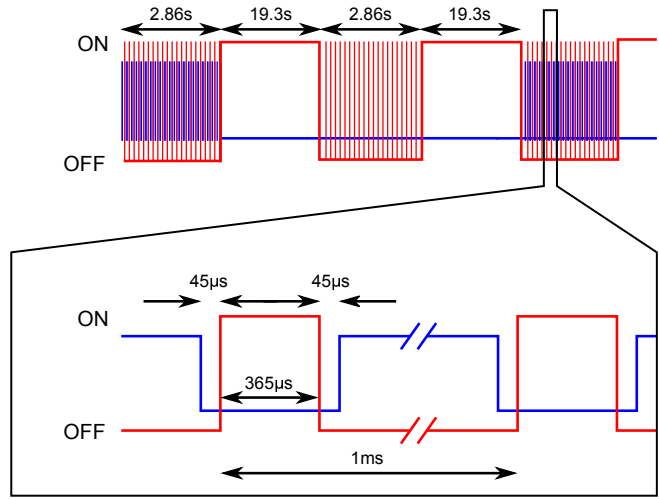}%
 \caption{Timing sequence of the ground-state photoionization cross-section measurement. Red and blue curve corresponds to the cooling and ionizing light, respectively. The inset shows the chopping phase in details.}\label{fig:timing}
 \end{figure}

The photoionization cross section $\sigma_{5S_{1/2}}^{253.7}$ was determined by comparing times of decaying of the MOT fluorescence with and without the presence of the photoionizing 253.7~nm light. To ensure all atoms were in the ground state during UV exposition, we used appropriate timing sequence of the experiment~\cite{Lowell2002,Ciampini2002}. After switching off all cooling and repumping beams by an acousto-optical modulator (AOM) we waited 45~$\mu$s before exposure to the UV which is much longer than the lifetime of the $5P_{3/2}$ state (27~ns). Subsequently, the atoms were illuminated by 275~$\mu$s long  pulse of the 253.7~nm light. The trapping beams were turned back on 45~$\mu$s after switching off the UV light. The cycle has been repeated until total depletion of atoms from the MOT. The total time of the sequence was equal to 2.86~s. Fig.~\ref{fig:timing} shows timing scheme in details. 

To switch the light within the shortest possible time, fast RF switches with 2~$\mu$s of switching time were used to drive the relevant AOMs. To ensure there are no stray resonant photons which may contribute to the excited-state fraction all possible sources of the resonant light were blocked during the UV operation. We checked also that no measurable leakage light was present on the AOM output.

To determine the photoionization cross section $\sigma_{5S_{1/2}}$ 212 decaying curves with chopped trapping and repumping beams were recorded with and without the alternating sequence of the UV pulses, giving decaying rates $\gamma_{P}^{253.7}+\gamma_{Rb}$ and $\gamma_{Rb}$, respectively. Afterwards, the averaged decaying rates $\langle\gamma_{P}^{253.7}+\gamma_{Rb}\rangle$ and $\langle\gamma_{Rb}\rangle$ were used to find $\sigma_{5S_{1/2}}^{253.7}$ according to the formula

\begin{equation}
  \sigma_{5S_{1/2}}^{253.7}=\frac{h\nu^{253.7}_P \left( \langle\gamma_{P}^{253.7}+\gamma_{Rb}\rangle - \langle\gamma_{Rb}\rangle \right)}{\langle{I}_{P}^{253.7}\rangle},\label{eq:sigma_ground}
\end{equation}

\noindent
where the average intensity of the photoionization light seen by the atoms $\langle{I}_{P}^{253.7}\rangle$ was calculated according to the equations (\ref{eq:avr_intensity}) -- (\ref{eq:N_distribution})

\begin{equation}
  \langle{I}_P^{253.7}\rangle=\frac{2PT\eta}{\pi \sqrt{\left(r_x^2+ \sigma_x^2\right)\left(r_y^2+ \sigma_y^2\right)}},
\end{equation}

\noindent
where $\eta$ is a duty cycle equals to 0.545, accordingly to the timing sequence.
We determined the photoionization cross section to be 

\begin{equation}
  \sigma_{5S_{1/2}}^{253.7}=1.0^{+4.3}_{-1}\times 10^{-20}\text{cm}^2.
\end{equation}

\noindent
The uncertainty of $\sigma_{5S_{1/2}}^{253.7}$ is the standard uncertainty under assumption that   $\sigma_{5S_{1/2}}^{253.7}$ cannot be negative.
It consists of two contributions, one related to the uncertainty of intensity of the photoionization light $\langle I_p^{253.7} \rangle$ and the second, due to the statistical spread of the differences between decaying rates. The statistical uncertainty is dominant because of low signal to noise ratio in measurements of the photoionization from the $\sigma_{5S_{1/2}}^{253.7}$ ground state. Surprisingly, even with the significant uncertainty, this result can be used for precise determination of the photoionization cross section of the excited state $\sigma_{5P_{3/2}}^{253.7}$.


\subsubsection{$5P_{3/2}$ state}
\begin{figure}[ht]
 \includegraphics[width=\columnwidth]{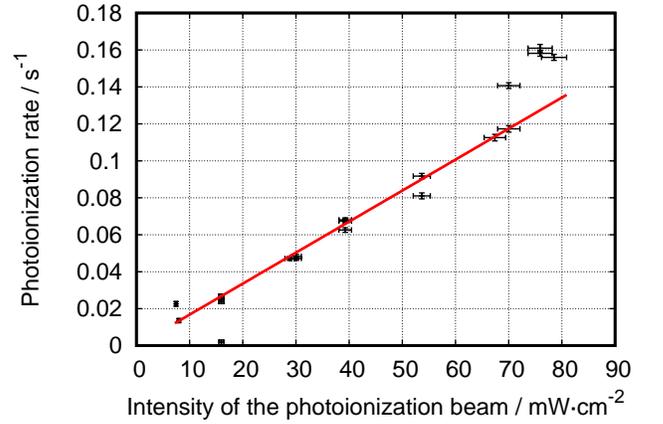}%
 \caption{Dependence of the 253.7~nm photoionization  rate $\gamma_P^{253.7}$ in ${}^{87}$Rb MOT on the photoionizing laser intensity $\langle I \rangle$.  The measurement was performed for the excited state fraction equals to 0.215(29). The linear fit (red solid line) to the formula~(\ref{eq:gamma253_}) gives the value of the $\sigma^{253.7}_{eff}$.}\label{fig:254sigma_eff}
\end{figure}

~
The determination of the photoionization cross section for the excited state $\sigma_{5P_{3/2}}^{253.7}$ was based on the formula~\eqref{eq:gamma253} which can be rewritten as 
\begin{equation}
  \gamma^{253.7}_P  =\frac{\sigma^{253.7}_{eff}\left( \rho_{5P_{3/2}} \right)}{h\nu^{253.7}_P}I^{253.7}_P \label{eq:gamma253_},
\end{equation}

\noindent
where 

\begin{equation}
  \sigma_{eff}^{253.7}\left( \rho_{5P_{3/2}} \right)=\left( \sigma^{253.7}_{5P_{3/2}}-\sigma^{253.7}_{5S_{1/2}}\right) \rho_{5P_{3/2}} + \sigma^{253.7}_{5S_{1/2}}\label{eq:sigma_vs_fraction}
\end{equation}

\noindent
is an effective photoionization cross section at 253.7~nm. The dependence of the $\sigma_{eff}^{253.7}$ on the excited-state fraction $\rho_{5P_{3/2}}$ has been measured for fractions from 0.215(29) to 0.357(18).
Each $\sigma_{eff}^{253.7}$ value was obtained from a linear fit according to the formula

\begin{equation}
  \sigma_{eff}^{253.7}=\frac{\gamma_{P}^{253.7}}{I_P^{253.7}}h\nu_{P}^{253.7},
\end{equation}

\noindent
where the photoionization rates $\gamma_P^{253.7}$ were determined by comparing exponential fits of the Rb MOT loading curves with 253.7~nm ionizing light on and off. The relative uncertainty  $u( \sigma_{eff}^{253.7})/\sigma_{eff}^{253.7}$ for each excited state fraction were estimated by linear regression to be less than 4\%. Fig.~\ref{fig:254sigma_eff} shows typical dependence of the photoionization rate  $\gamma_P^{253.7}$ on the photoionizing laser intensity measured for the excited state fraction of 0.215(29).

\begin{figure}[ht]
  \includegraphics[width=\columnwidth]{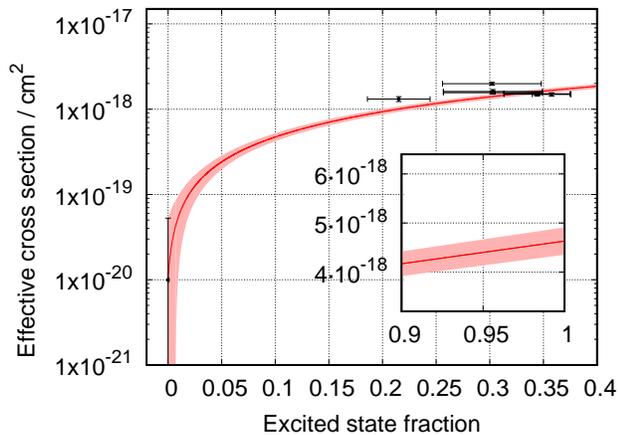}
  \caption{Effective ionization cross section 
    $\sigma_{eff}^{253.7}$ versus excited state fraction $\rho_{5P_{3/2}}$. The solid red curve is a weighted linear fit to the data. The transparent red area corresponds to the statistical uncertainty of the fit. To improve visibility of results the logarithmic scale was used. The inset shows the area nearby the fully populated excited state, i.e. $\rho_{5P_{3/2}}=1$. 
}
  \label{fig:sigma_vs_fraction} 
\end{figure}

Fig.~\ref{fig:sigma_vs_fraction} shows the dependence of the effective cross section $\sigma_{eff}^{253.7}$ on the excited state fraction $\rho_{5P_{3/2}}$.
The determined above $\sigma^{253.7}_{5S_{1/2}}$ corresponds to the $\rho_{5P_{3/2}}=0$.
The uncertainties of the excited state fraction have been calculated as the standard deviation of the weighted mean of the data measured at a given intensity of the MOT beams.
The photoionization cross section of the excited state $\sigma^{253.7}_{5P_{3/2}}$ has been determined from a weighted linear fit (red line in Fig.~\ref{fig:sigma_vs_fraction}) of the formula~(\ref{eq:sigma_vs_fraction}) with $\sigma^{253.7}_{5P_{3/2}}$ being the free parameter.
The fit yields a value of 

\begin{equation}
  \sigma^{253.7}_{5P_{3/2}}=4.63(30)\times 10^{-18}\text{cm}^2.
\end{equation}

\noindent
The uncertainty of $\sigma^{253.7}_{5P_{3/2}}$ involves the accuracy of the weighted linear fit as well as the uncertainty of the measurement of $\sigma^{253.7}_{5S_{1/2}}$.
~
 

\begin{figure}[ht]
  \includegraphics[width=\columnwidth]{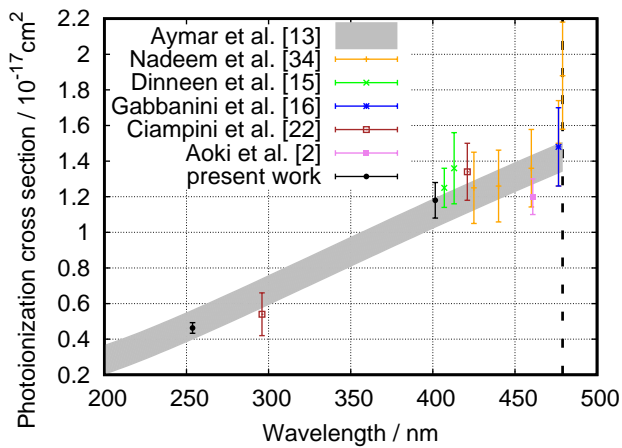}
  \caption{Photoionization cross section of $5P_{3/2}$~Rb state. Grey area corresponds to the theoretical results inferred from plot. The vertical dashed line denotes the photoionization treshold. Filled circles are our results. Other data points are experimental results extracted from literature.}
  \label{fig:5P_comparison} 
\end{figure}

\begin{figure}[ht]
  \includegraphics[width=\columnwidth]{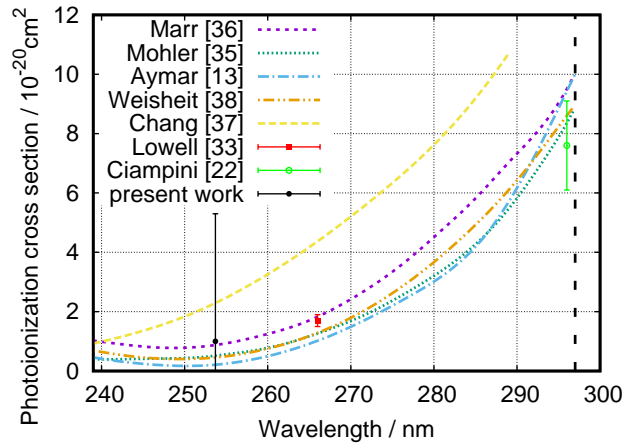}
  \caption{Photoionization cross section of $5S_{1/2}$~Rb state. The vertical dashed line denotes the photoionization treshold. Filled circle is our result. Other data points are experimental (Mohler, Marr, Lowell, Ciampini), theoretical (Chang, Aymar) and semiempirical (Weisheit) results extracted from literature. }
  \label{fig:5S_comparison} 
\end{figure}
\section{Discussion}

Our results are compared with data determined by other groups. Figures 9 and 10 show the dependence of the photoionization cross sections for $5P_{3/2}$ and $5S_{1/2}$ states on laser wavelength, respectively.  

All our results are consistent with previously reported experimental results~\cite{Dinneen1992,Gabbanini1997,Ciampini2002,Nadeem2011,Aoki2013,Mohler1929,Marr1968,Lowell2002} as well as theoretical predictions~\cite{Aymar1984,Chang1972,Weisheit1972}. The value of $\sigma^{253.7}_{5S_{1/2}}$ determined here confirms earlier results indicating that near this wavelength the Cooper minimum of the photoionization cross section for $5S_{1/2}$ state is expected~\cite{Aymar1984,Mohler1929,Marr1968,Weisheit1972}. Our measured value $\sigma^{253.7}_{5P_{3/2}}$ is the first experimental determination of the photoionization cross section at this wavelength. Moreover, at this point it is the most precise determination of the photoionization cross section for $5P_{3/2}$ state.

\noindent
\section{Conclusion}
In conclusion, we have determined the values of the Rb photoionization cross sections at 253.7~nm to be $1.0^{+4.3}_{-1}\times10^{-20}\text {cm}^2$ and  $4.63(30)\times 10^{-18}\text{cm}^2$ for $5S_{1/2}$ and $5P_{3/2}$ states, respectively. Additionally, the photoionization cross section for $5P_{3/2}$ state at 401.5~nm was found to be $1.18(10)\times 10^{-17}~\text{cm}^{2}$.
Our results are consistent with the data reported by other groups.
We have found out that the cross section for the photoionization of Rb close to 254~nm is small enough to restrain the Rb losses in the case when the trap is spatially overlapped with Hg MOT.

\section*{Acknowledgements}

We thank Marcin Bober for providing us with 401.5~nm laser light.
The reported measurements have been performed in the
National Laboratory FAMO in Toru\'n and supported by the
subsidy of the Ministry of Science and Higher Education. MZ contribution is supported by the National Science Centre, Poland, Project no. 2017/25/B/ST2/00429. P\.Z, MW and RMR contribution is supported by the National Science Centre, Poland, Project no. 2012/07/B/ST2/00235. RC and BN contribution is supported by the National Science Centre, Poland, Project no. 2013/11/N/ST2/00846.

\nocite{Chang1972}
\nocite{Marr1968}
\nocite{Mohler1929}
\nocite{Weisheit1972}
\bibliography{hgrb_photoionization}

\end{document}